\documentclass{SciPost}

\hypersetup{
    colorlinks,
    linkcolor={red!50!black},
    citecolor={blue!50!black},
    urlcolor={blue!80!black}
}

\newcommand{\magenta}[1]{\color{magenta} #1 \color{black}}

\usepackage[bitstream-charter]{mathdesign}
\usepackage{url}
\usepackage{color}
\usepackage{fontawesome}

\DeclareSymbolFont{usualmathcal}{OMS}{cmsy}{m}{n}
\DeclareSymbolFontAlphabet{\mathcal}{usualmathcal}

\fancypagestyle{SPstyle}{
\fancyhf{}
\lhead{\colorbox{scipostblue}{\bf \color{white} ~SciPost Physics Community Reports }}
\rhead{{\bf \color{scipostdeepblue} ~Submission }}

\fancyfoot[C]{\textbf{\thepage}}
}

\begin{document}\vspace*{-1cm}
\begin{flushleft} 
\magenta{LHCHWG-2026-008}\\
\magenta{IFT-UAM/CSIC-26-84}
\end{flushleft}

\pagestyle{SPstyle}

\begin{center}{\Large \textbf{\color{scipostdeepblue}{
Impact of NLO EW and QCD corrections to dimension-6 SMEFT coefficients constraints in Higgs decays\\
}}}\end{center}

\begin{center}\textbf{
Luigi Bellafronte\textsuperscript{1$\star$},
Ana Rosario Cueto Gómez \textsuperscript{2$\dagger$},
Sally Dawson\textsuperscript{3$\ddagger$},
Clara Del Pio\textsuperscript{3$\circ$},\\
Matthew Forslund\textsuperscript{4$\S$},
Pier Paolo Giardino\textsuperscript{5$\P$} and 
Andrea Visibile \textsuperscript{6$\dagger\dagger$}
}\end{center}

\begin{center}
{\bf 1} Physics Department, Florida State University, Tallahassee, FL 32306-4350, USA
\\
{\bf 2} Departamento de Física Teórica, 
    Universidad Autónoma de Madrid, Cantoblanco, 28049, Madrid, Spain
\\    
{\bf 3} Physics Department, 
    Brookhaven National Laboratory, Upton, NY 11973, USA
\\
{\bf 4} Princeton Center for Theoretical Science, Princeton University, Princeton, NJ, 08544 USA
\\
{\bf 5} Departamento de Física Teórica and Instituto de Física Teórica UAM/CSIC, 
    Universidad Autónoma de Madrid, Cantoblanco, 28049, Madrid, Spain
\\    
{\bf 6} Stockholm University, SE-106 91 Stockholm, Sweden    
\\[\baselineskip]
$\star$ \href{mailto:lbellafronte@fsu.edu}{\small lbellafronte@fsu.edu}\,,\quad
$\dagger$ \href{mailto:anar.cueto@uam.es}{\small anar.cueto@uam.es}\,,\quad
$\ddagger$ \href{mailto:dawson@bnl.gov}{\small dawson@bnl.gov}\,,\quad
$\circ$ \href{mailto:cdelpio@bnl.gov}{\small cdelpio@bnl.gov}\,, \\
$\S$ \href{mailto:mforslund@princeton.edu}{\small mforslund@princeton.edu}\,,\quad
$\P$ \href{mailto:pier.giardino@uam.es}{\small pier.giardino@uam.es}\,, \quad 
$\dagger\dagger$ \href{mailto:andrea.visibile@cern.ch}{\small andrea.visibile@cern.ch}\,, \quad 
\end{center}

\section*{\color{scipostdeepblue}{Abstract}}
\textbf{\boldmath{%
This paper presents the impact of the inclusion of next-to-leading-order (NLO) QCD and electroweak (EW) corrections for dimension-6 Standard Model Effective Field Theory (SMEFT) predictions for the Higgs decays in Higgs couplings measurements at the Large Hadron Collider. Such higher-order corrections will be increasingly important for the precise Higgs and electroweak measurements expected at the High-Luminosity Large Hadron Collider (HL-LHC). The results are compared with a SMEFT parameterisation employing inputs from a previous ATLAS study, in which the dimension-6 SMEFT contributions are evaluated for each production and decay process at the lowest non-vanishing perturbative QCD order of the corresponding SM process. The effects of the SMEFT NLO QCD/EW contributions to the Higgs decay channels are largest for 2-fermion operators involving the third generation. The sensitivity to the Higgs tri-linear coupling is also explored.}
}

\vspace{\baselineskip}

\noindent\textcolor{white!90!black}{%
\fbox{\parbox{0.975\linewidth}{%
\textcolor{white!40!black}{\begin{tabular}{lr}%
  \begin{minipage}{0.6\textwidth}%
    {\small Copyright attribution to authors. \newline
    This work is a submission to SciPost Physics. \newline
    License information to appear upon publication. \newline
    Publication information to appear upon publication.}
  \end{minipage} & \begin{minipage}{0.4\textwidth}
    {\small Received Date \newline Accepted Date \newline Published Date}%
  \end{minipage}
\end{tabular}}
}}
}


\vspace{10pt}
\noindent\rule{\textwidth}{1pt}
\tableofcontents
\noindent\rule{\textwidth}{1pt}
\vspace{10pt}

\section{Introduction}
\label{sec:intro}
The absence so far of direct evidence for new particles at the LHC has made effective field theory an increasingly central tool for indirect searches for physics beyond the Standard Model (SM). In the dimension-six Standard Model Effective Field Theory (SMEFT), deviations from SM predictions induced by new physics at a scale $\Lambda$ are parameterised by a set of higher-dimensional operators, whose Wilson coefficients can be constrained from precision measurements without reference to a specific ultraviolet completion~\cite{Brivio:2017vri}.

Within this framework, interpretations of combined Higgs boson measurements have been carried out by both the ATLAS and CMS Collaborations~\cite{ATLAS:2024lyh,CMS:2026nce}, exploiting the information contained in the different Higgs production modes and decay channels to constrain the Wilson coefficients of the operators entering the Higgs sector. In essentially all such analyses, the dependence of the Higgs production and decay rates on the Wilson coefficients is extracted at leading order (LO). A notable exception concerns loop-induced production modes and decay channels: $gg\to H$, $gg\to ZH$, $H\to \gamma\gamma$,  $H\to Z\gamma$ and $H\to gg$, for which the SM amplitude itself first appears at one loop and the SMEFT parameterisation includes next-to-leading order quantum chromodynamics (QCD) or electroweak (EW) corrections. For all other channels, however, the parameterisation used in the experimental fits remains strictly tree-level.

While QCD corrections to predictions in the dimension-6 SMEFT are well-known and have been implemented in automated codes both for Higgs production and decay, in recent years, the calculation of  the NLO  electroweak corrections in the dimension-six SMEFT has progressed substantially. The complete set of NLO QCD and EW corrections is now available for all 2- and 4-body Higgs decays in the narrow-width approximation~\cite{Bellafronte:2025jbk,Bellafronte:2026jic}. These calculations show that NLO corrections, and in particular electroweak corrections,  can modify the relative weight of different Wilson coefficients entering the Higgs decay rates, raising the question of whether constraints derived from a purely LO parameterisation remain a reliable guide. Furthermore, at NLO, new Wilson coefficients contribute that are not present at tree level. The complete QCD and EW NLO corrections to Higgs decays are now available using the POPfx format~\cite{Brivio:2025mww} to aid experimental and phenomenological studies. The purpose of this paper is to examine the numerical consequences of the LO parameterization used in current fits  and to demonstrate the potential importance of a full NLO fit including both SM and SMEFT contributions at NLO.

In view of the higher statistics available in the High Luminosity phase of the LHC, the question of the effects induced by NLO corrections in the SMEFT on the extraction of the bounds on New Physics is becoming more and more pressing. Moreover, full NLO results can enhance the constraining power on those operators that appear at loop level, either from the direct contribution to loop diagrams or by operator mixing in the renormalization group equations. In this paper, we use the theoretical results from~\cite{Bellafronte:2025jbk,Bellafronte:2026jic} to quantify the impact of the NLO QCD/EW contributions to Higgs decay,  as compared to the LO parameterization so far employed by the experimental collaborations. We perform single parameter fits, neglecting potential correlations between different SMEFT operators. The analysis is performed using LHC Run 2 Higgs measurements, corresponding to an integrated luminosity of 140~${\rm fb}^{-1}$. While the resulting constraints on the Wilson coefficients are expected to be weaker than those attainable at the HL-LHC, the relative impact of employing the LO versus NLO parameterization  of SMEFT effects is expected to remain of similar magnitude. This type of analysis quantifying the relative impact of NLO vs LO SMEFT limits, including all Higgs decay channels, is novel to this work. Starting from the Run 2 combined Higgs measurements from ATLAS~\cite{ATLAS:2024lyh} and the updated SMEFT setup as used in~\cite{ATLAS:2026fyh}, we compare the constraints on the relevant Wilson coefficients obtained using the LO parameterization of the Higgs decay rates against those obtained when the NLO QCD and EW corrections are included, and assess the impact of this choice on the derived limits. The study also shows those operators appearing only through higher-order corrections, for which sensitivity is already achieved with Run 2 measurements, exposing the need to include NLO SMEFT corrections for HL-LHC interpretations.

\section{Methodology}
The dimension-6 SMEFT Lagrangian is written as,
\begin{equation}
{\cal {L}}={\cal{L}}_{SM}+\Sigma_i{c_i\over \Lambda^2} {\cal{O}}_i^{(6)}\, ,
\end{equation}
where $\mathcal{L}_{SM}$ is the pure Standard Model Lagrangian and ${\cal{O}}_i^{(6)}$ is the complete set of $SU(3)\times SU(2)\times U(1)$ gauge invariant dimension-six operators\cite{Grzadkowski:2010es}.  With this convention, the Wilson coefficients $c_i$ are dimensionless.  In our numerical results, we choose $\Lambda=1$ TeV.

The constraints on the Wilson coefficients are derived from a multivariate Gaussian likelihood, built from the covariance matrices of the simplified-template cross-section (STXS) measurements used in the ATLAS Higgs interpretations~\cite{ATLAS:2024lyh,ATLAS:2026fyh}. The likelihood $L\left(\boldsymbol x | \boldsymbol{c}\right)$ for an individual measurement is modelled as a multivariate Gaussian:
\begin{equation}
  \begin{aligned}
    L\left(\boldsymbol x \middle| \boldsymbol{c}\right)=&\frac{1}{\sqrt{\left(2\pi\right)^{n_\text{bins}}\mathrm{det}\left(V\right)}}\exp\left(-\frac{1}{2}\Delta {\boldsymbol x}^{\intercal}\left(\boldsymbol{c}\right) V^{-1}\Delta{\boldsymbol x}\left(\boldsymbol{c}\right)\right),
    \end{aligned}
    \label{eq:likelihood}
\end{equation}
\noindent
 where $V$ is the covariance matrix containing the information on the statistical and systematic uncertainties of the measurement, $\boldsymbol{c}$ is the vector of Wilson coefficients, the vector 
 \begin{equation}     \Delta{\boldsymbol x (\boldsymbol{c})}=(\Delta x_1,\dots, \Delta x_{n_\text{bins}})
 \end{equation}
 is the difference between measurement and prediction for a given set of Wilson coefficients and $n_{bins}$ is the number of bins. This Gaussian approximation of the full STXS likelihood was validated in those analyses against the corresponding full likelihood function, which retains the complete set of nuisance parameters and non-Gaussian effects of the original measurements; the differences between the two were found to be small \footnote{ See Appendix of~\cite{ATLAS:2024lyh}.}, supporting the use of the covariance-matrix-based approach adopted here.

The dependence of the Higgs boson production cross-sections on the Wilson coefficients is taken unchanged from~\cite{ATLAS:2026fyh}. For the majority of production processes, this dependence is computed at LO using MadGraph5\_aMC@NLO~\cite{Alwall:2014hca} interfaced with the SMEFTsim 3.0 UFO model~\cite{Brivio:2020onw}. Loop-induced processes are instead generated with SMEFT@NLO~\cite{Degrande:2020evl}. Both models employ the  Warsaw basis~\cite{Grzadkowski:2010es}. SMEFTsim notation is used throughout with operators restricted to the "top" flavour scheme, which imposes a $U(2)^3$ flavour symmetry in the quark sector and a $U(1)^3$ flavour symmetry in the lepton sector. The parameterisation described here is kept fixed for this study.

The Higgs boson partial and total decay widths are instead taken from the dedicated NLO QCD and electroweak SMEFT calculation of~\cite{Bellafronte:2026jic}, which provides results for all two- and four-body Higgs decays, using the Fermi constant $G_\mu$ and the masses of the $W$ and $Z$ bosons as inputs. The Wilson coefficients are renormalized in the $\overline{\rm MS}$ scheme, while the  gauge couplings and all masses are renormalized on-shell. This choice ensures consistency with the input-parameter and renormalization scheme used in the production-mode parameterization. To isolate the impact of the higher-order corrections, the decay rates from~\cite{Bellafronte:2026jic} are used twice in otherwise identical fits: once truncated at LO, and once including the full NLO QCD and electroweak corrections truncated at ${\cal{O}}({1\over\Lambda^2})$. The parameterization of the production mode times branching ratio is also linearized in the Wilson coefficients. It was checked that the LO parameterization of tree-level decay channels coincide with the parameterization used previously by the experimental collaborations. This study does not consider potential differences in the measurement acceptances caused by new physics, as such an analysis is beyond the scope of this work. Comparing the Wilson coefficient constraints obtained in these two configurations (LO and NLO SMEFT analyses), with the statistical framework and production parameterization held fixed as described above, directly quantifies the impact of the NLO Higgs decay corrections on the SMEFT constraints from combined Higgs measurements.

\section{Results}

\begin{figure*}[t]
        \centering
\includegraphics[width=0.99\textwidth]{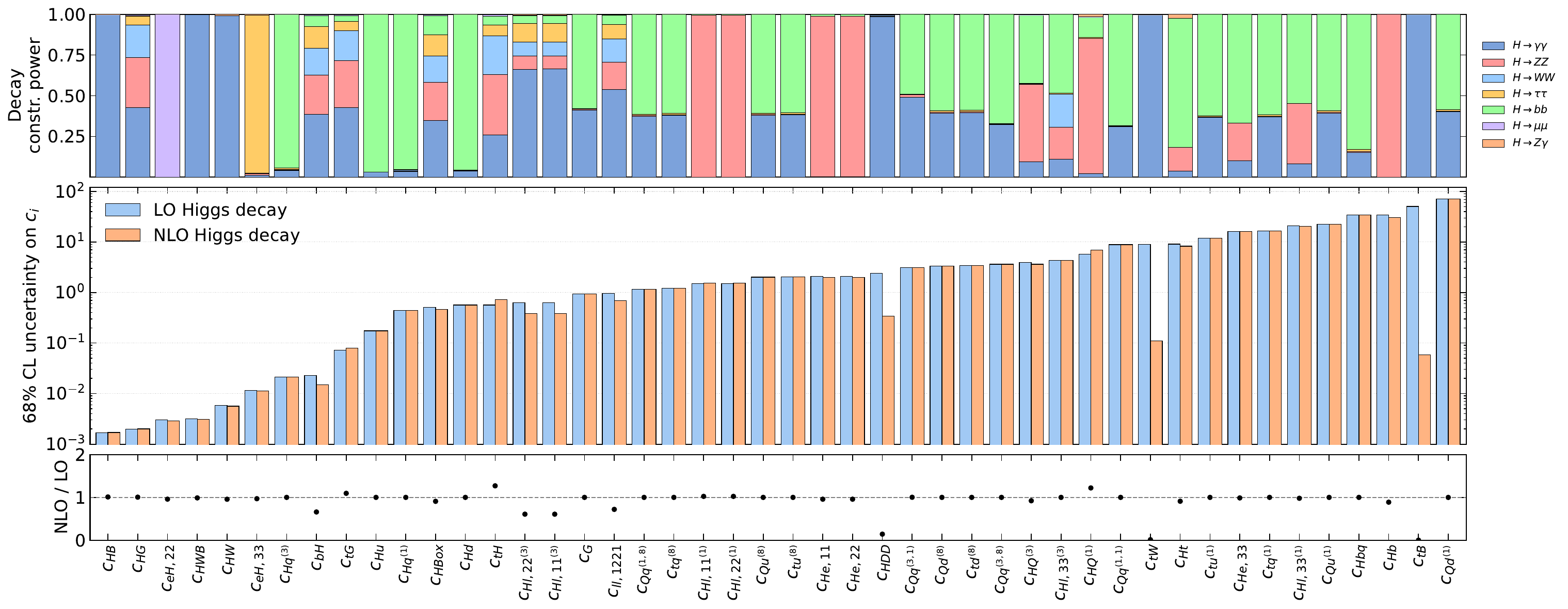}
        \caption{68\% CL one-parameter expected uncertainties on Warsaw basis Wilson coefficients assuming either a LO or NLO parameterization of the Higgs Branching ratios. The upper plot shows the relative constraining power of each of the Higgs boson decays measurements for each of the operators. Note that the Higgs production contributions in the dimension-6 SMEFT are included at NLO QCD for loop-induced processes and at LO EW for tree-level ones. The cut-off scale $\Lambda$ was set to $1$~TeV.}
        \label{fg:LO_vs_NLO}
\end{figure*}

\begin{figure*}[t]
        \centering
\includegraphics[width=0.99\textwidth]{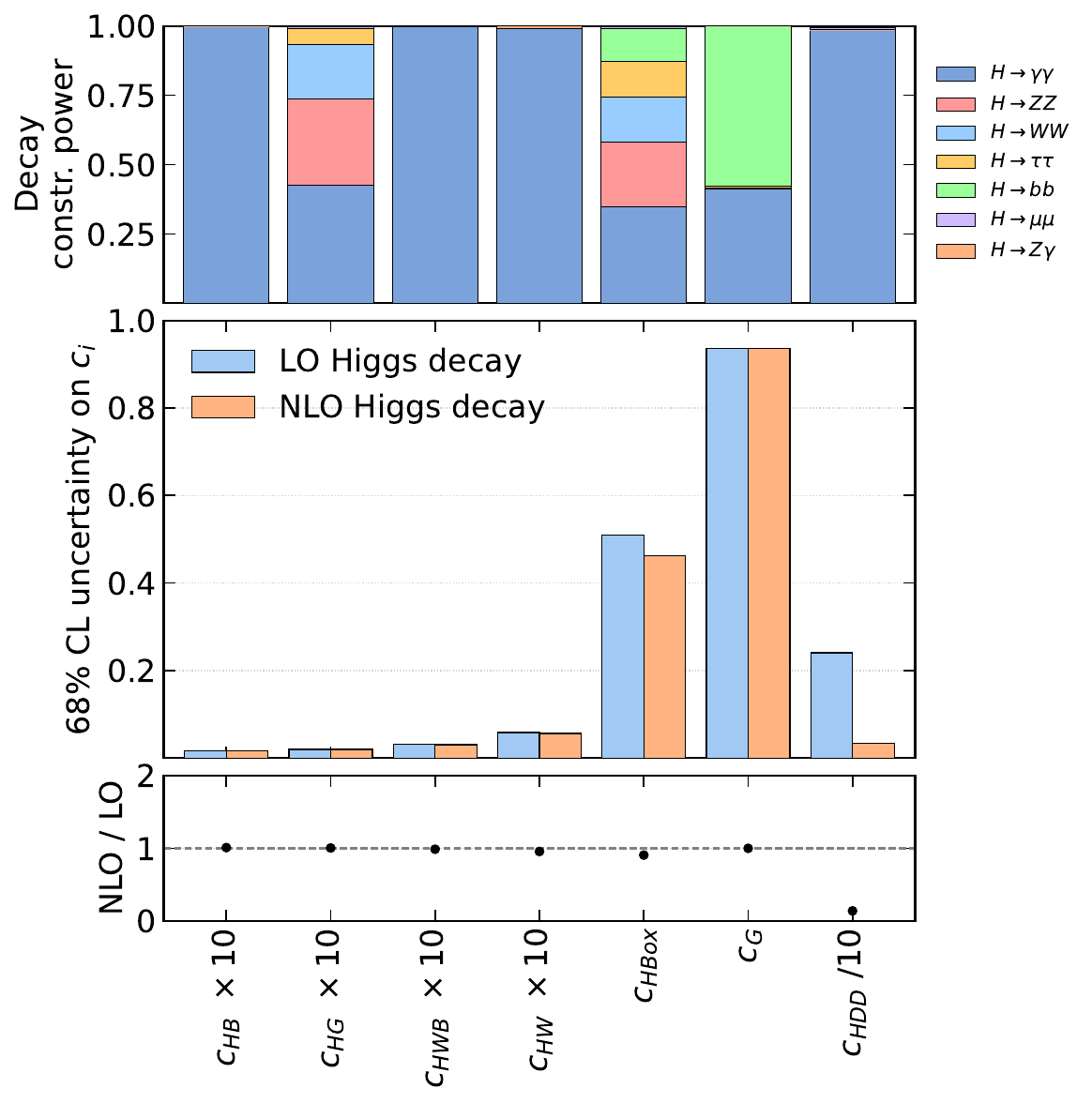}
        \caption{68\% CL one-parameter expected uncertainties on Warsaw basis bosonic Wilson coefficients assuming either a LO or NLO parameterization of the Higgs branching ratios. Some constraints are scaled by the factors indicated in the labels of their operators to improve their visualization. The upper plot shows the relative constraining power of each of the Higgs boson decays measurements for each of the operators. Note that the Higgs production contributions in the dimension-6 SMEFT are included at NLO QCD for loop-induced processes and at LO EW for tree-level ones. The cut-off scale $\Lambda$ was set to $1$~TeV.}
        \label{fg:LO_vs_NLO_bosonic}
\end{figure*}

\begin{figure*}[t]
        \centering
\includegraphics[width=0.99\textwidth]{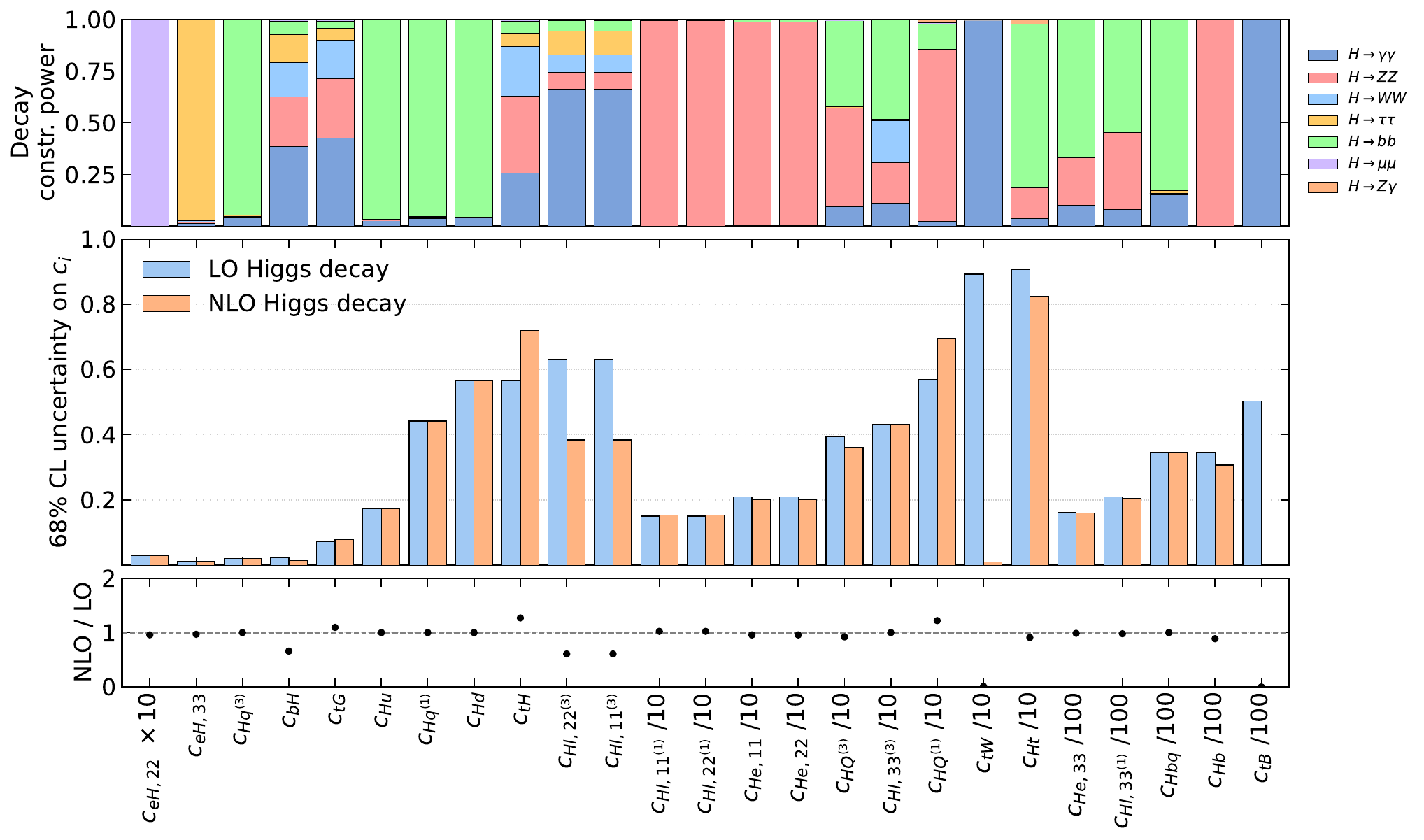}
        \caption{68\% CL one-parameter expected uncertainties on Warsaw basis two-fermion Wilson coefficients assuming either a LO or NLO parameterization of the Higgs branching ratios. Some constraints are scaled by the factors indicated in the labels of their operators to improve their visualization. The upper plot shows the relative constraining power of each of the Higgs boson decays measurements for each of the operators. Note that the Higgs production contributions in the dimension-6 SMEFT are included at NLO QCD for loop-induced processes and at LO EW for tree-level ones. The cut-off scale $\Lambda$ was set to $1$~TeV.}
        \label{fg:LO_vs_NLO_twofermion}
\end{figure*}

\begin{figure*}[t]
        \centering
\includegraphics[width=0.99\textwidth]{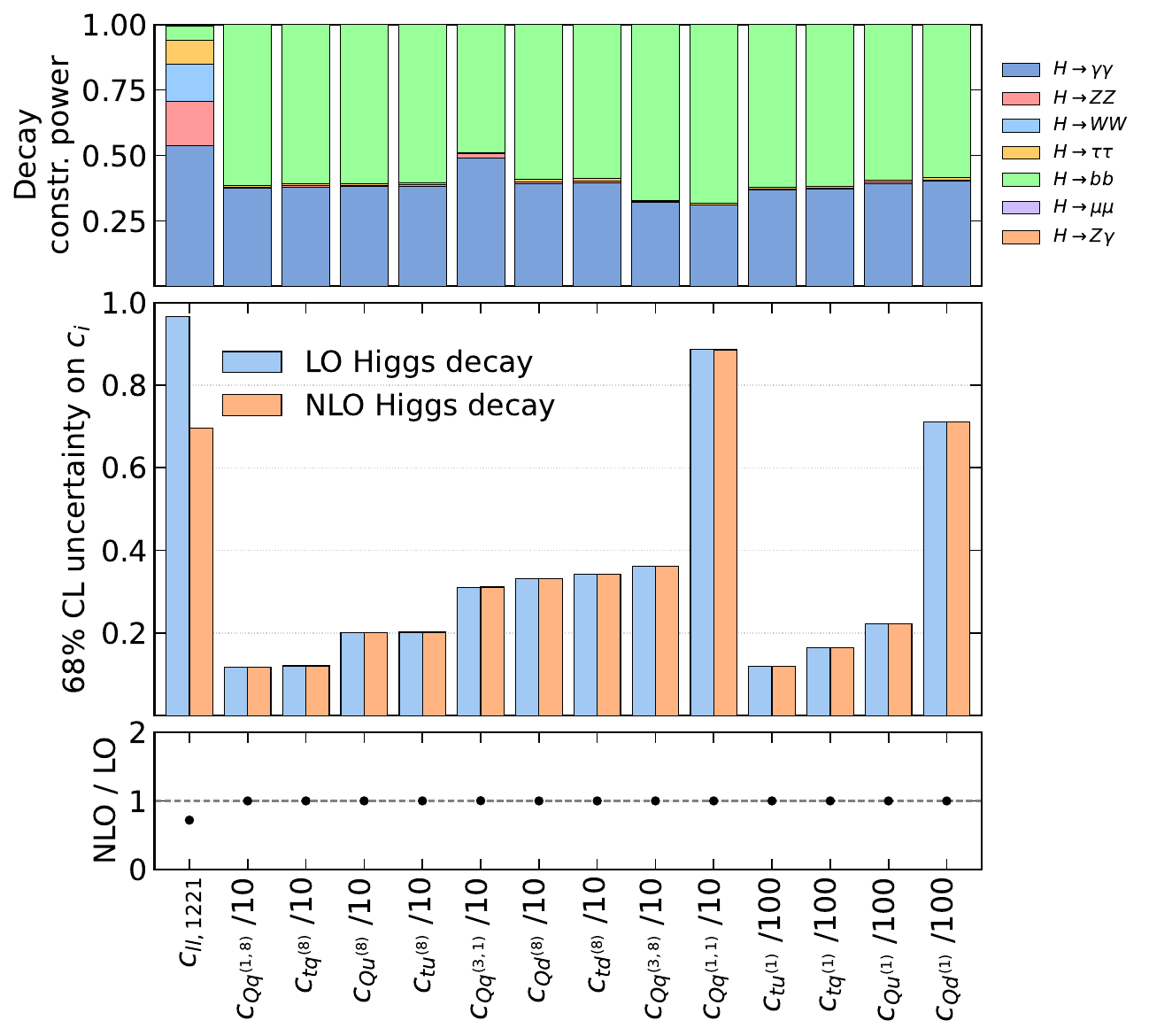}
        \caption{68\% CL one-parameter expected uncertainties on Warsaw basis four-fermion Wilson coefficients assuming either a LO or NLO parameterization of the Higgs Branching ratios. Some constraints are scaled by the factors indicated in the labels of their operators to improve their visualization. The upper plot shows the relative constraining power of each of the Higgs boson decays measurements for each of the operators. Note that the Higgs production contributions in the dimension-6 SMEFT are included at NLO QCD for loop-induced processes and at LO EW for tree-level ones. The cut-off scale $\Lambda$ was set to $1$~TeV.}
        \label{fg:LO_vs_NLO_fourfermion}
\end{figure*}

{Figure~\ref{fg:LO_vs_NLO} provides an overview of the impact of including NLO QCD and electroweak corrections to the Higgs decay parameterisation on the SMEFT fit. It presents the 68\% CL one-parameter expected uncertainties on the Wilson coefficients already constrained at LO, comparing the results obtained with the LO and NLO decay widths from Ref.~\cite{Bellafronte:2026jic}, while keeping the production-mode parameterisation fixed throughout, in order to isolate the effect of the higher-order corrections in the decay sector. The upper panel illustrates the relative contribution of each Higgs decay channel measurement to the total constraining power on a given operator, showing how the sensitivity is distributed across the $H\to\gamma\gamma$, $H\to ZZ^*$, $H\to WW^*$, $H\to\tau\tau$, $H\to b\bar{b}$, $H\to\mu\mu$, and $H\to Z\gamma$ channels. To facilitate the discussion, the Wilson coefficients are grouped according to their operator class in the Warsaw basis, with the results for purely bosonic operators, operators involving two fermionic and two bosonic fields, and four-fermion operators presented separately in Figures~\ref{fg:LO_vs_NLO_bosonic}, \ref{fg:LO_vs_NLO_twofermion}, and \ref{fg:LO_vs_NLO_fourfermion}, respectively.}
 \footnote{We use the notation of \cite{Brivio:2020onw,ATLAS:2026fyh} to define the Wilson coefficients, which corresponds to the SMEFTsim notation.  To convert to the  familiar notation of \cite{Dedes:2017zog}, take $H\rightarrow \phi$ in the definitions of the coefficient functions and $c_{HDD}\rightarrow c_{\phi D}$.}

For a large number of operators, the NLO/LO ratio of the 68\% CL expected uncertainties, shown in the lower panel of Figure~\ref{fg:LO_vs_NLO}, lies close to unity, indicating that the inclusion of NLO QCD and electroweak corrections in the decay widths does not significantly shift the extracted constraints for those operators contributing at tree-level. This major result of this study is consistent with the expectation that, for operators whose tree-level contribution to the decay rates is non-vanishing, the NLO corrections are subleading at the level of precision currently achieved by the data. Nevertheless, shifts at the level of 20--30\% are observed for a subset of operators in the intermediate-sensitivity range, namely $c_{bH}$, $c_{tH}$, $c_{Hl,22}^{(3)}$, $c_{Hl,11}^{(3)}$, $c_{HQ,1}^{(3)}$, and $c_{Hb}$. These operators receive comparatively large electroweak corrections to fermionic Higgs decay amplitudes, making their extracted constraints more sensitive to higher-order effects than those of the remaining tree-level operators. The $c_{Hl,22}^{(3)}$ and $c_{Hl,11}^{(3)}$ operators constraints are typically also affected by the measurement acceptance of the new physics which is not explicitly studied here. Even larger sensitivities are found for the operators $c_{HDD}$, $c_{tW}$, and $c_{tB}$, which contribute significantly to the NLO electroweak corrections of loop-induced decay channels, in particular $H\to\gamma\gamma$. The LO versus NLO shifts for a given operator are attributable solely to the change in the decay-rate parameterization. These relative effects are expected to remain of the same magnitude for the HL-LHC phase, while small changes could happen if the relative constraining power of the different measurements change or new measurements enter the combination. These one-operator-at-a-time constraints do not speak to how correlations among operators in a global fit would respond to the NLO update.

\begin{figure*}[t]
        \centering
\includegraphics[width=0.99\textwidth]{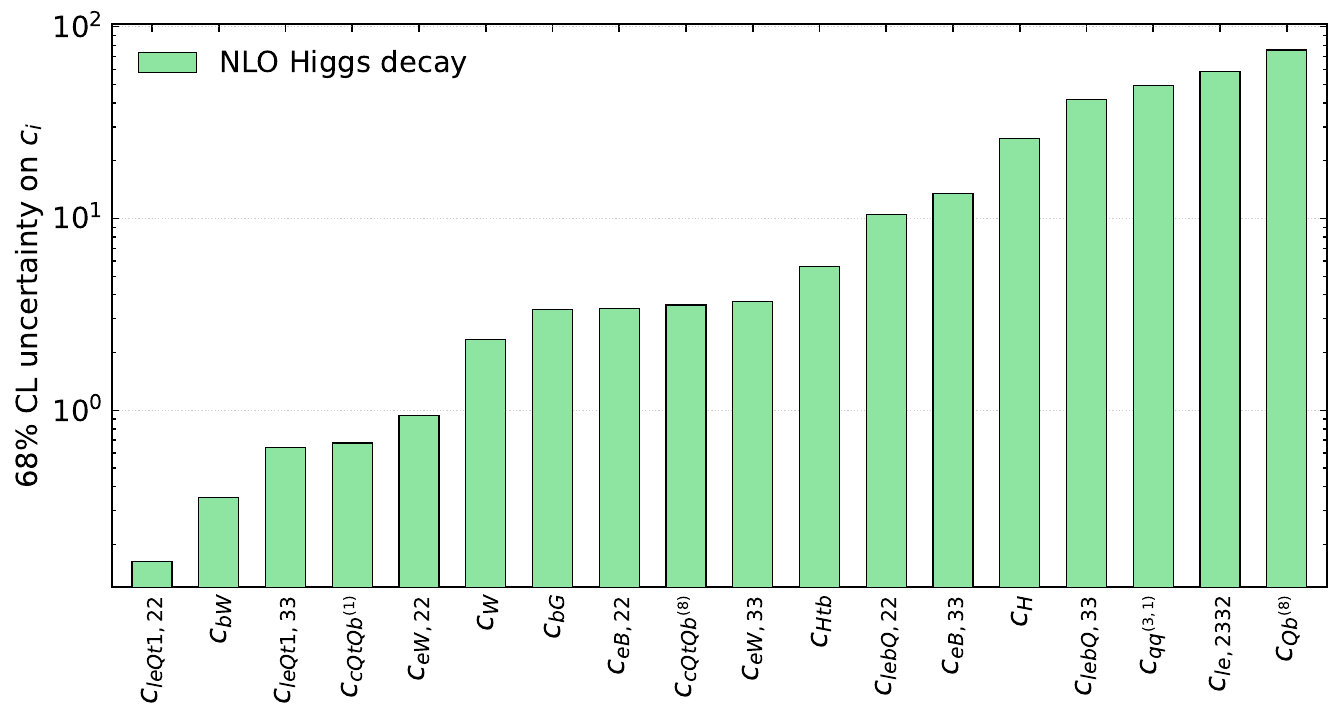}
        \caption{68\% CL one-parameter expected uncertainties on Warsaw basis Wilson coefficients that appear only through higher-order SMEFT corrections. Only operators with uncertainties below 100 are shown in the plot. The cut-off scale $\Lambda$ was set to $1$~TeV.}
        \label{fg:NLO_only}
\end{figure*}

The inclusion of higher-order corrections in the SMEFT brings new sensitivity to dimension-six operators that do not contribute to  the lowest order rates for the  QCD Higgs production modes or to the decays at LO. Figure~\ref{fg:NLO_only} shows the 68\% CL one-parameter uncertainties on this new set of operators, obtained from the NLO decay parameterization of the Higgs decay widths.
It shows how stringent constraints can be obtained on operators such as $c_{leQt1,22}$, $c_{leQt1,33}$, $c_{cQtQb}^{(1)}$, $c_{bW}$ and $c_{eW,22}$  which will only improve with more precise HL-LHC measurements. Notably, the $c_H$ operator, which modifies the Higgs self-coupling, enters the fit for the first time through the inclusion of NLO EW decay contributions in the linearized SMEFT. This allows experimental collaborations to constrain further this operator using single-Higgs measurements. The most stringent constraints on $c_H$ from Higgs decays come from the $H\rightarrow \gamma\gamma$ and $H\rightarrow ZZ^*$ channels.

\clearpage

\section{Conclusions}
 The effects of NLO QCD and EW corrections to 2- and 4- body Higgs decays in the dimension-6 SMEFT were studied in the context of an ATLAS SMEFT analysis that included both production and decay channels  at LO in the SMEFT. Including the decays at NLO  QCD and EW order in  the SMEFT quantifies the impact of the NLO dimension-6 SMEFT corrections to Higgs decays.  In general, the effects of the NLO corrections to the Higgs decays are small for operators contributing at tree-level, with the exception of some operators involving the third-generation fermions where effects of the ${\cal{O}}(20-30\%) $ are seen.  Significant constraints are  obtained on coefficients that contribute to Higgs decays only at NLO in the dimension-six  SMEFT. The relative improvement of the constraints is expected to be of the same magnitude in the HL-LHC phase, while the additional sensitivity to operators that contribute to Higgs decays only at NLO makes these effects increasingly relevant for future SMEFT interpretations.

A global fit to LHC data that includes the NLO QCD/EW SMEFT contributions to the production processes requires the 2-loop SMEFT contribution to the $gg\rightarrow H$ process, along with the NLO SMEFT contributions to the vector boson scattering processes and  to $t {\overline{t}}H $ production.  These calculations are difficult, but technically feasible and can be expected to be completed in the next few years.  The Higgstrahlung process at NLO QCD/EW SMEFT is also needed, and is in progress.

\section*{Acknowledgements}
S.D. and C.D.P. are supported
by the U.S. Department of Energy under Contract No. DE-
SC0012704.  
P.P.G. is supported by the Ramón y Cajal grant~RYC2022-038517-I funded by MCIN/AEI/10.13039/501100011033 and by FSE+, and by the R\&D\&I Project CEX2025-001574-S, funded by MICIU/AEI/10.13039/501100011033 within the Particle Physics in the Standard Model and Beyond research line. The work of L.B. is supported in part by the U.S.
Department of Energy under Grant No. DE-SC0010102 and by the College of Arts and Sciences of Florida State University. A.R.C.G is supported by the Ramón y Cajal grant~RYC2021-031273-I funded by MICIU/AEI
/10.13039/501100011033 and the European Union NextGenerationEU/PRTR, the PID2024-156748NB-I00 project with additional funding from FEDER/UE, and the European Research Council (ERC) under the European Union’s Horizon Europe research and innovation programme (grant agreement No 101219398). A.V work is supported by the Knut and Alice Wallenberg foundation under the grant KAW 2023.0366. A.V also acknowledges financial support from Vetenskapsrådet under Grant No. 2022-04981.

\bibliography{references.bib}

\end{document}